\documentclass[%
  reprint,
  superscriptaddress,
  amsmath,amssymb,
  prb,
]{revtex4-2}

\usepackage{graphicx}
\usepackage{dcolumn}
\usepackage{bm}
\usepackage{hyperref}
\usepackage{tikz}
\usepackage{epigraph}
\usetikzlibrary{decorations.pathmorphing,positioning,arrows.meta}
\hypersetup{
  colorlinks=true,
  linkcolor=blue,
  citecolor=blue,
  filecolor=blue,
  urlcolor=blue,
}
\usepackage{hypcap}
\newcommand{\commented}[1]{}
\usepackage[dvipsnames]{xcolor} 
\begin{document}

\preprint{APS/123-QED}


%
%
\title{Temperature-dependent vibrational EELS simulations with nuclear quantum effects}

%
%
\author{Zuxian He}
\affiliation{Department of Physics and Astronomy, Uppsala University, Box 516, 75120 Uppsala, Sweden}
\author{J\'an Rusz}
\email{jan.rusz@physics.uu.se}
\affiliation{Department of Physics and Astronomy, Uppsala University, Box 516, 75120 Uppsala, Sweden}

\date{\today}
%
%
\begin{abstract}
  The Time Autocorrelation of Auxiliary Wave (TACAW) method has established a framework for modeling angle-resolved electron energy loss spectroscopy (EELS) of phonons and magnons by deriving scattering intensities from the time autocorrelation of the beam wavefunction. This approach enables efficient computation of scattering intensities while naturally accounting for dynamical diffraction and multiple-scattering effects. In the cryogenic regime, vibrational spectra are dominated by nuclear quantum effects, notably zero-point motion. To capture these effects in low-temperature vibrational EELS, we incorporate thermostatted ring polymer molecular dynamics (TRPMD) into the TACAW formalism. Our results demonstrate that nuclear quantum effects lead to significant deviations from classical molecular dynamics predictions in the vibrational spectra of silicon at low temperatures and correctly predict the nearly temperature-independent optical phonon peak intensities in silicon, consistent with the first Born approximation. The TRPMD-TACAW method provides a robust theoretical tool for probing the low-temperature limit of vibrational EELS, offering a necessary benchmark for the quantitative analysis of emerging cryogenic scanning transmission electron microscopy experiments.
\end{abstract}
\maketitle

%
%
\section{Introduction}

Vibrational electron energy-loss spectroscopy (EELS) in the scanning transmission electron microscope (STEM) has emerged as a powerful probe of lattice dynamics \cite{krivanek2014nature}. Advances in electron monochromation, detector sensitivity, and beam stability have made it possible for the technique to provide direct access to phonon excitations in solids, allowing for the investigation of vibrational properties with unprecedented spatial and energy resolution. In the past decade, vibrational EELS has been applied to a wide range of materials problems, including the mapping of localized phonon modes \cite{hage_single-atom_2020,yan2024nanoscale,Yang2024,han2026electron}, the study of lattice dynamics at interfaces and defects \cite{yan_single-defect_2021,haas2023atomic,li2025single}, isotope effects \cite{hachtel_identification_2019,senga_imaging_2022,li2023phonon}, and the local thermometry \cite{IdroboPhysRevLett.120.095901, lagos2018thermometry, KikkawaPhysRevB.106.195431}. Many of these studies have focused on room-temperature or elevated-temperature conditions, where nuclear motion is reasonably approximated by classical dynamics.

Recent instrumental developments are extending vibrational EELS into the cryogenic regime. In particular, the implementation of liquid-helium cryostages for monochromated STEM has enabled stable measurements at temperatures below 10 K \cite{johnson2025high}. At such low temperatures, lattice dynamics are increasingly dominated by quantum nuclear effects, including zero-point motion and quantum fluctuations of the phonon, which cannot be captured within a purely classical description of nuclear trajectories \cite{lofgren_influence_2016,kim2024modeling}. Access to this regime is of particular interest for the study of quantum materials, where low-temperature lattice dynamics often play a central role in phase transitions and emergent collective phenomena such as superconductivity. Consequently, a quantitative theoretical framework that accounts for nuclear quantum effects becomes essential for the interpretation of cryogenic vibrational EELS experiments. 

Within the TACAW framework, angle-resolved energy-loss spectra are computed while retaining dynamical diffraction, multiple scattering, and the full multislice description of beam propagation \cite{tacawpaper}. In this approach, the scattering intensity is obtained from the time autocorrelation of an auxiliary beam wave function. This formulation renders the method computationally efficient and naturally suited for large-scale simulations.
Existing implementations \cite{tacawpaper,walker2026pyslice}, however, have relied on classical molecular dynamics to generate the underlying nuclear trajectories. This approximation becomes increasingly restrictive at low temperatures, where zero-point motion and quantum delocalization can significantly alter both structural fluctuations and vibrational spectra.

Path-integral methods provide a controlled description of equilibrium quantum nuclei by mapping the quantum partition function onto an isomorphic classical ring polymer. Within this framework, TRPMD combines exact quantum Boltzmann sampling with an approximate dynamical scheme that has been successfully applied to a wide range of condensed-phase systems, capturing nuclear quantum effects in both structural and dynamical properties \cite{markland2018nuclear, althorpe2024path,ceriotti2016nuclear,marsalek2017quantum,habershon2013ring}. These features make TRPMD a natural extension of the TACAW framework in the regime where quantum nuclear effects become significant.

In this work, we formulate and implement the TRPMD–TACAW approach to study the temperature-dependent vibrational EELS. We apply the method to crystalline silicon to investigate how nuclear quantum effects influence vibrational scattering across a wide temperature range. Our results show that the TRPMD and classical molecular dynamics descriptions become nearly indistinguishable at high temperatures, where thermal fluctuations dominate nuclear motion. In contrast, clear deviations emerge in phonon EELS as the temperature is lowered into the cryogenic regime, reflecting the increasing importance of quantum nuclear fluctuations. Notably, with the TRPMD-TACAW framework, we reproduce the nearly temperature-independent intensity of the optical phonon peak, as is expected on the basis of the Born approximation \cite{nicholls_theory_2019}. The combined TRPMD–TACAW framework, therefore, provides a quantitative model for vibrational EELS in regimes where classical descriptions of nuclear motion are no longer sufficient.

The remainder of the paper is organized as follows. Section~\ref{sec:theory} reviews the path-integral formulation and TRPMD used to describe nuclear quantum effects. Since these approaches are not commonly used in electron microscopy, a summary of the essential theoretical framework is reviewed in order to make the Article more self-contained. Section~\ref{sec:TRPMD_TACAW} introduces the TRPMD-TACAW extension of the TACAW method. Section~\ref{sec:results} presents the results for equilibrium ring-polymer observables such as mean squared displacements, velocity auto-correlation functions, and phonon EELS. Section~\ref{sec:conclusion} concludes the paper.

\section{Theory} \label{sec:theory}
In this section, we briefly review the ring-polymer formulation that underlies ring-polymer molecular dynamics (RPMD) \cite{craig2004quantum}. The central idea is to rewrite the quantum canonical partition function as that of an isomorphic classical system consisting of $n$ replicas (beads) of each particle connected by harmonic springs. This mapping makes it possible to sample quantum Boltzmann statistics using classical-like trajectories in an extended phase space. Within this framework, we also discuss several dynamical observables relevant to the present study, including the mean squared displacement, the radius of gyration of the ring polymer, and the velocity autocorrelation function. The corresponding time-correlation functions are evaluated using the Kubo-transformed correlation formalism, which provides the appropriate quantum statistical definition for dynamical quantities in path-integral molecular dynamics.

\subsection{Thermostatted ring-polymer molecular dynamics}

Consider a system of $N$ distinguishable particles with the Hamiltonian
\begin{equation}
  \begin{aligned}
    \hat{H}&=\hat{T}+\hat{V}\\
    &=\frac{1}{2}\hat{\mathbf{p}}^{\mathsf{T}}\mathbf{M}^{-1}\hat{\mathbf{p}}+V\left(\hat{\mathbf{q}}\right),
  \end{aligned}
\end{equation}
where $\hat{\mathbf{q}}=(\hat{\mathbf{q}}_1,\ldots,\hat{\mathbf{q}}_N)$ and $\hat{\mathbf{p}}=(\hat{\mathbf{p}}_1,\ldots,\hat{\mathbf{p}}_N)$ collect the position and momentum operators, and $\mathbf{M}=\mathrm{diag}(m_1,\ldots,m_N)$ is the mass matrix. The canonical partition function at inverse temperature $\beta=\left(k_B T\right)^{-1}$ is
\begin{equation}
  Z=\operatorname{Tr}\left(e^{-\beta \hat{H}}\right).
\end{equation}

Using the symmetric Trotter factorization \cite{trotter1959product}
\begin{equation}
e^{-\beta \hat{H}}=\lim _{n \rightarrow \infty}\left(e^{-\beta_n \hat{V} / 2} e^{-\beta_n \hat{T}} e^{-\beta_n \hat{V} / 2}\right)^n,
\end{equation}
with $\beta_n=\beta / n$, the partition function can be written as a discretized imaginary-time path integral, and $n$ is the Trotter number, corresponding to the number of imaginary-time slices. Inserting complete sets of position eigenstates between the factors yields \cite{10.1063/1.441588}

\begin{widetext}
  \begin{equation}\label{eq:partitonfZn}
    \begin{aligned}
      Z_n=&\int d \mathbf{q}_1 \cdots d \mathbf{q}_n \prod_{j=1}^n\left\langle \mathbf{q}_j\right| e^{-\frac{\beta_n}{2} \hat{V}} e^{-\beta_n \hat{T}} e^{-\frac{\beta_n}{2} \hat{V}}\left|\mathbf{q}_{j+1}\right\rangle\\
      =&\left[\prod_{a=1}^{N}\left(\frac{m_a}{2 \pi \hbar^2 \beta_n}\right)^{n / 2}\right]
      \int \prod_{j=1}^{n} d \mathbf{q}_j\,
      \exp\Bigg[-\beta_n \sum_{j=1}^{n}\Bigg(V\left(\mathbf{q}_{1, j}, \ldots, \mathbf{q}_{N, j}\right)+\sum_{a=1}^{N} \frac{1}{2} m_a \omega_n^2\left(\mathbf{q}_{a,j}-\mathbf{q}_{a,j+1}\right)^2\Bigg)\Bigg],
    \end{aligned}
  \end{equation}
\end{widetext}
here $\omega_n=1 /\left(\beta_n \hbar\right)$, and cyclic boundary conditions $\mathbf{q}_{a, n+1} \equiv \mathbf{q}_{a, 1}$ are imposed.

By introducing auxiliary momenta allows the configurational integral to be written in classical phase-space form,
\begin{equation}
Z_n \propto \int d \mathbf{p} d \mathbf{q} e^{-\beta_n H_n(\mathbf{p}, \mathbf{q})},
\end{equation}
where the effective ring-polymer Hamiltonian is

\begin{equation}
  \begin{aligned}
    H_n(\mathbf p,\mathbf q)=
    &  \sum_{i=1}^N\sum_{j=1}^n
      \left[
        \frac{p_{i,j}^2}{2m'_i}
        +
        \frac{1}{2} m_i \omega_n^2
        \left(q_{i,j}-q_{i,j-1}\right)^2
    \right]\\
    &\quad+
    \sum_{j=1}^n V(\mathbf q_j).
  \end{aligned}
  \label{eq:Hrp_main}
\end{equation}

Where $m^{\prime}_a$ is a fictitious mass, and for static observables in the PIMD formulation, the results are independent of its choice. For dynamic observables calculated by RPMD the situation is different (see below). The Hamiltonian $H_n(\mathbf{p}, \mathbf{q})$ describes a classical system composed of $n$ coupled replicas of each particle, forming a closed ring polymer. Each bead experiences the physical potential $V$, while neighboring beads are linked by harmonic springs with frequency $\omega_n$. In the limit $n \rightarrow \infty$, the discretized representation converges to the exact imaginary-time path-integral formulation of the quantum partition function.

In practical simulations, the number of imaginary-time slices must be large enough to resolve the fastest intrinsic time scale of the system. If the system has a maximum characteristic frequency $\omega_{\text {max }}$, the shortest relevant time scale is on the order of $\omega_{\max }^{-1}$. Since the imaginary-time interval $[0, \beta \hbar]$ is discretized into steps of size $\Delta \tau=\beta \hbar / n$, accuracy requires $\Delta \tau \ll \omega_{\max }^{-1}$. This leads to the practical criterion
\begin{equation}\label{eq:boundbeads}
  n \gtrsim \beta \hbar \omega_{\text {max }},
\end{equation}
which ensures that high-frequency quantum fluctuations are properly resolved and that Trotter discretization errors remain controlled \cite{markland2008efficient}. In practice, convergence is verified by increasing $n$ until thermodynamic observables become insensitive to further refinement.


The equilibrium ring-polymer distribution defined by the Hamiltonian $H_n(\mathbf{p}, \mathbf{q})$ can be sampled by microcanonical dynamics in the extended phase space. In practice, however, the harmonic spring interactions between beads introduce a set of high-frequency internal modes, whose frequencies increase with the number of beads. These stiff modes lead to poor ergodicity and slow convergence of equilibrium averages \cite{hall1984nonergodicity}.

To improve sampling efficiency, thermostatted ring-polymer molecular dynamics (TRPMD) introduces stochastic coupling to a thermal bath. A simple and widely used choice for canonical sampling is the Langevin thermostat, which couples stochastic friction and noise terms to the equations of motion and enforces the correct Boltzmann distribution. However, when applied directly in Cartesian bead coordinates, the Langevin thermostat may introduce spurious resonances between the thermostat and the high-frequency normal modes of the ring polymer. These artificial resonances (see Appendix~\ref{sec:diffThermostat}) can distort the dynamics and reduce sampling efficiency \cite{rossi2014remove}.

In this work, we employ the path-integral Langevin equation (PILE) thermostat, in which Langevin dynamics is applied in the normal-mode representation of the ring polymer \cite{ceriotti2010efficient}. In these coordinates, the free ring-polymer Hamiltonian becomes a set of independent harmonic oscillators with frequencies
\begin{equation}
  \omega_k=2 \omega_n \sin \left(\frac{k \pi}{n}\right).
\end{equation}
The dynamics of each normal mode $k$ are governed by the Langevin equations for a particle is 
\begin{equation}
    \small
\dot{\tilde{q}}_j^{(k)}=\frac{\tilde{p}_j^{(k)}}{m_j}, \quad \dot{\tilde{p}}_j^{(k)}=-m_j \omega_k^2 \tilde{q}_j^{(k)}-\gamma_k \tilde{p}_j^{(k)}+\sqrt{\frac{2 m_j \gamma_k}{\beta_n}} \xi_j^{(k)}(t),
\end{equation}
where $\xi_i^{(k)}(t)$ denotes Gaussian white noise. The friction coefficients $\gamma_k$ are chosen to optimize sampling efficiency. For the internal ring-polymer modes $(k>0)$, minimizing the autocorrelation time of the harmonic oscillator energy yields the optimal choice $\gamma_k=2 \omega_k$. This strong thermostatting rapidly equilibrates the fictitious internal modes without affecting physical observables. The centroid mode $k=0$, which contains the physically meaningful motion of the system, is instead coupled only weakly to the bath using a small friction $\gamma_0=\tau_0^{-1}$.

Two commonly used variants differ in how the centroid is thermostatted. In PILE-L, the centroid momentum components are coupled independently to Langevin baths, resulting in local stochastic dissipation. In PILE-G, the thermostat acts globally on the total centroid kinetic energy through stochastic velocity rescaling. The choice of thermostat in TRPMD impacts the resulting EELS spectra, see Appendix \ref{sec:diffThermostat}.

\subsection{Dynamical observables from TPRMD}\label{subsec:ObsTRPMD}
At low temperatures, nuclear quantum effects become significant and are well approximated within the ring-polymer formalism. TRPMD allows these effects to be observed through simple dynamical observables. In the following, we introduce the squared radius of gyration, the mean squared displacement, and the velocity autocorrelation function, which is closely related to the vibrational density of states. These quantities highlight the crossover from classical to quantum-dominated behavior.

The squared radius of gyration of a ring polymer consisting of $N$ distinguishable particles and $n$ beads per particle is defined as
\begin{equation}
  \begin{aligned}
    R_G^2(t_{k})=&\frac{1}{N n} \sum_{i=1}^N \sum_{j=1}^n\left|\vec{r}_{i, j}(t_{k})-\vec{R}_i(t_{k})\right|^2\\
    =&\frac{1}{N n} \sum_{i=1}^N \sum_{j=1}^n\left|\delta \vec{r}_{i, j}(t_{k})\right|^2,
  \end{aligned}
\end{equation}
where
\begin{equation}
  \delta \vec{r}_{i, j}(t_{k}):=\vec{r}_{i, j}(t_{k})-\vec{R}_i(t_{k}), \quad \vec{R}_i(t_{k})=\frac{1}{n} \sum_{i=1}^n \vec{r}_{i,n}(t_{k}),
\end{equation}
and $\vec{R}_n(t_{k})$ describes the centroid motion. The squared radius of gyration provides a measure of the quantum delocalization encoded in the imaginary-time path-integral representation. Its value depends sensitively on temperature: in the high temperature limit ($T\gg 0$), the harmonic springs connecting the beads become stiff, the polymer collapses toward its centroid, and $R_G^2 \rightarrow 0$, recovering the classical point-particle limit. In contrast, at low temperatures, the springs soften, and the ring polymer spreads out, leading to a finite radius of gyration that reflects the zero-point motion of the underlying quantum particles. This temperature dependence makes $R_G$ a useful diagnostic of quantum effects in path-integral molecular dynamics.

A second important dynamical observable is the ring-polymer mean squared displacement (MSD), defined as
\begin{equation}
\small
\begin{aligned}
   \text{MSD}& _{\text{TRPMD}}(t_{k}) \\
    =& \frac{1}{N n} \sum_{i=1}^N \sum_{j=1}^n\lvert \vec{r}_{i,j}(t_{k})-\vec{r}_{i,j}(t_{0}) \lvert^{2}\\
    =&\frac{1}{N n} \sum_{i=1}^{N} \sum_{j=1}^{n} \lvert (\vec{R}_{i}(t_{k})-\vec{R}_{i}(t_{0}))+(\delta \vec{r}_{i,j}(t_{k})-\delta \vec{r}_{i,j}(t_{0})) \lvert^{2}\\
    =&\frac{1}{N n} \sum_{i=1}^{N} \sum_{j=1}^{n}(\lvert \vec{R}_{i}(t_{k})-\vec{R}_{i}(t_{0}) \lvert^{2}+\lvert \delta \vec{r}_{i,j}(t_{k})-\delta \vec{r}_{i,j}(t_{0}) \lvert^{2}).
\end{aligned}
\end{equation}
Here $t_{0}$ is taken as an arbitrary reference frame. If one picks $t_{0}$ to be the static equilibrium, we find
\begin{equation}\label{eq:MSDRPMD}
  \text{MSD} _{\text{TRPMD}}(t_{k})= \text{MSD}_{\text{centroid}}(t_{k})+R_G^2(t_{k}).
\end{equation}
In the classical limit, where $n=1$, each particle is represented by a single bead, and the ring polymer carries no internal fluctuations. Thus, the radius of gyration vanishes, and the TRPMD mean squared displacement reduces to the classical one.
\begin{equation}
  \operatorname{MSD}_{\mathrm{TRPMD}}\left(t_k\right)=\operatorname{MSD}_{\text {centroid }}\left(t_k\right) \quad(n=1
  ).
\end{equation}
The same behavior arises at high temperatures. The inter-bead springs stiffen, the ring polymer collapses onto its centroid, and quantum delocalization disappears. In this limit $R_G^2 \rightarrow 0$, the TRPMD trajectory again coincides with the centroid trajectory, yielding the same result as in the $n=1$ case,
\begin{equation}
  \operatorname{MSD}_{\mathrm{TRPMD}}\left(t_k\right)=\operatorname{MSD}_{\text {centroid }}\left(t_k\right) \quad(T \gg 0 \text{ K}).
\end{equation}

Finally, in the low-temperature limit, the centroid effectively becomes frozen; thus, its motions vanish, and the MSD is determined entirely by the quantum delocalization within each ring polymer. In this regime, we have
\begin{equation}
  \operatorname{MSD}_{\mathrm{TRPMD}}\left(t_k\right)=R_G^2\left(t_k\right)\quad(T \approx 0 \text{ K}).
\end{equation}

In summary, Eq.~\eqref{eq:MSDRPMD} shows that the TRPMD mean squared displacement naturally separates into a centroid contribution, describing the classical-like motion, and a purely quantum contribution $R_G^2$ that accounts for the internal delocalization of the ring polymer. At low temperatures, where centroid motion can be strongly suppressed, $R_G^2$ therefore provides the dominant correction to the classical picture.

A third dynamical observable of central importance is the velocity autocorrelation function (VACF), defined as
\begin{equation}
\langle v(\tau) \cdot v(0)\rangle=\frac{1}{N} \sum_{i=1}^N v_i(0+\tau) \cdot v_i(0),
\end{equation}
where $v(\tau)$ is the centroid velocity of the ring-polymer. The Fourier transform of VACF (FVACF) provides direct access to the vibrational density of states and characteristic phonon frequencies, making it a convenient diagnostic for assessing how nuclear quantum effects modify lattice dynamics.

\subsection{Correlation functions}\label{subsec:TRPMD_corr}

Real-time quantum correlation functions provide access to the dynamical and spectral properties of many-body systems, but are generally difficult to compute exactly. Within the path-integral framework, static equilibrium averages are readily accessible, while approximate schemes are required for time-dependent correlations. In this subsection, we introduce the ring-polymer estimators for thermal expectation values and correlation functions and briefly discuss their relation to the Kubo-transformed quantum correlation function.

For two operators depending only on nuclear coordinates, $\hat{A}=A(\hat{x})$ and $\hat{B}=B(\hat{x})$, the exact quantum real-time correlation function is defined as
\begin{equation}
C_{A B}(t)=\frac{1}{Z} \operatorname{Tr}\left[e^{-\beta \hat{H}} \hat{A} e^{+i \hat{H} t / \hbar} \hat{B} e^{-i \hat{H} t / \hbar}\right] .
\end{equation}
In the ring-polymer representation, one introduces bead-averaged estimators \cite{craig2004quantum}
\begin{equation}
A_n(\mathbf{x})=\frac{1}{n} \sum_{j=1}^n A\left(x_j\right), \quad B_n(\mathbf{x})=\frac{1}{n} \sum_{j=1}^n B\left(x_j\right),
\end{equation}
and considers the corresponding equilibrium phase-space average
\begin{equation}
\langle A B\rangle_n=\frac{1}{(2 \pi \hbar)^n Z_n} \int d \mathbf{p} d \mathbf{x} e^{-\beta_n H_n(\mathbf{p}, \mathbf{x})} A_n(\mathbf{x}) B_n(\mathbf{x})
\end{equation}
In the limit $n \rightarrow \infty$, this quantity does not correspond to the thermal expectation value of the operator product $\hat{A} \hat{B}$. Instead, it yields the $t \rightarrow 0$ limit of the Kubo-transformed correlation function \cite{kubo_statistical_1957,craig2004quantum}
\begin{equation}
  \small
  \widetilde{C}_{A B}(t)=\frac{1}{\beta Z} \int_0^\beta d \lambda \operatorname{Tr}\left[e^{-(\beta-\lambda) \hat{H}} \hat{A} e^{-\lambda \hat{H}} e^{+i \hat{H} t / \hbar} \hat{B} e^{-i \hat{H} t / \hbar}\right].
\end{equation}
RPMD proceeds by propagating the ring-polymer phase-space variables using the classical equations of motion generated by the ring-polymer Hamiltonian $H_n$. Here in the ring-polymer Hamiltonian $H_n$, the fictitious bead masses are to be considered the physical masses of the particles. Starting from initial conditions $(\mathbf{p}_0, \mathbf{x}_0)$ sampled from the equilibrium distribution $e^{-\beta_n H_n}$, the trajectories $\mathbf{x}_t=\mathbf{x}_t\left(\mathbf{p}_0, \mathbf{x}_0\right)$ generate the approximate time-correlation function
\begin{equation}
\langle A(0) B(t)\rangle_n= \int d \mathbf{p}_0 d \mathbf{x}_0 \frac{e^{-\beta_n H_n\left(\mathbf{p}_0, \mathbf{x}_0\right)}}{(2 \pi \hbar)^n Z_n} A_n\left(\mathbf{x}_0\right) B_n\left(\mathbf{x}_t\right).
\end{equation}
This expression provides the RPMD approximation to the Kubo-transformed correlation function $\widetilde{C}_{A B}(t)$. It is exact at $t=0$ and reproduces the correct short-time limit of the quantum dynamics.

The spectrum obtained from the Kubo-transformed correlation function is related to the exact quantum spectrum through the following exact relation \cite{kubo_statistical_1957}
\begin{equation}\label{eq:kuboTransf}
  G_{A B}(\omega)=\frac{\beta \hbar \omega}{1-e^{-\beta \hbar \omega}} \widetilde{G}_{A B}(\omega).
\end{equation}
where $G_{A B}(\omega)$ and $\widetilde{G}_{A B}(\omega)$ denote the Fourier transforms of $C_{A B}(t)$ and $\widetilde{C}_{A B}(t)$, respectively. In the classical limit $\beta \hbar \omega \ll 1$, the two spectra coincide. More generally, the Kubo-transformed correlation function provides an exact description for systems with harmonic potentials or when the operators $A$ and $B$ are linear in position and momentum operators. Under these conditions, the resulting spectrum coincides with the exact quantum result, whereas deviations may arise for strongly anharmonic potentials or nonlinear operators \cite{craig2004quantum}.

\section{TRPMD-TACAW formalism}\label{sec:TRPMD_TACAW}
Recently, a new computational method has been introduced for simulating angle-resolved electron energy-loss (EELS) and energy-gain (EEGS) spectroscopies of phonons and magnons in transmission electron microscopy. Central to this approach is the time auto-correlation of the auxiliary wavefunction (TACAW), which enables the direct extraction of fractional scattering intensities from the temperature-dependent temporal evolution of the electron beam \cite{tacawpaper}. This framework naturally captures single and multiple scattering processes, dynamical diffraction effects, and explicit temperature dependence. Moreover, the method is computationally efficient and well suited for large-scale parallelization and GPU acceleration.

The TACAW method expresses the angle- and energy-resolved intensity, $I(\boldsymbol{q},E)$, where $\boldsymbol{q}$ is the transverse wavevector in the detector plane and $E$ is the energy transfer between the initial and final states, as the Fourier transform of the quantum-mechanical time autocorrelation function of the beam transmission operator $\hat{\phi}$. Specifically,
\begin{equation}
  I(\boldsymbol{q}, E)=\int_{-\infty}^{\infty} \frac{e^{-\frac{i}{\hbar} E t}}{2 \pi \hbar} c_{\phi \phi}(t) d t=c_{\phi \phi}(E),
\end{equation}
where
\begin{equation}
  c_{\phi \phi}(t)=\langle\hat{\phi}(t) \hat{\phi}(0)\rangle=\frac{1}{Z} \operatorname{Tr}\left[e^{-\beta \hat{H}_c} \hat{\phi}^{\dagger}(\boldsymbol{q}, 0) \hat{\phi}(\boldsymbol{q}, t)\right],
\end{equation}
and $\hat{\phi}(\boldsymbol{q}, t)=e^{i \hat{H}_c t / \hbar} \hat{\phi}(\boldsymbol{q}, 0) e^{-i \hat{H}_c t / \hbar}$. In a direct quantum treatment, this correlation function is prohibitively expensive for realistic many-body systems.

The TACAW implementation replaces the exact correlation function by a classical trajectory average. Denoting the classical estimator by $\tilde{c}_{\phi\phi}(t)$, the quantum spectrum is approximated through the Kubo relation
\begin{equation}
  c_{\phi \phi}(E)=\frac{\beta E}{1-e^{-\beta E}} \tilde{c}_{\phi \phi}(E),
\end{equation}
where $\tilde{c}_{\phi \phi}(E)$ is the Fourier transform of the classical correlation function. In molecular-dynamics practice, the spectrum is obtained from the Fourier transform of the time-dependent auxiliary wave,
\begin{equation}
  \tilde{c}_{\phi \phi}(E) \propto\left|\phi_{\mathrm{MD}}(\boldsymbol{q}, E)\right|^2,
\end{equation}
where 
\begin{equation}
\phi_{\mathrm{MD}}(\boldsymbol{q}, E)=\int d t e^{-i E t / \hbar} \phi(\boldsymbol{q}, t),
\end{equation}
which gives
\begin{equation}
  I(\boldsymbol{q}, E) \propto \frac{\beta E}{1-e^{-\beta E}}\left|\phi_{M D}(\boldsymbol{q}, E)\right|^2.
\end{equation}

The TRPMD extension follows directly from the ring-polymer representation of the thermal ensemble, Section~\ref{subsec:TRPMD_corr}. We replace the classical auxiliary wave with the bead-averaged estimator
\begin{equation}
  \phi_n(\mathbf{q}, t)=\frac{1}{n} \sum_{j=1}^n \phi\left(\boldsymbol{q} ; \mathbf{R}^{(j)}(t)\right),
\end{equation}
where $\mathbf{R}^{(j)}(t)$ denotes the nuclear coordinates of bead $j$ at time $t$. The corresponding TRPMD-TACAW spectrum is then
\begin{equation}
  I(\boldsymbol{q}, E) \propto \frac{\beta E}{1-e^{-\beta E}}\left|\phi_\text{TRPMD}(\boldsymbol{q}, E)\right|^2,
\end{equation}
with
\begin{equation}
  \phi_{\mathrm{TPRMD}}(\boldsymbol{q}, E)=\int d t \, e^{-i E t / \hbar} \phi_{n}(\boldsymbol{q}, t).
\end{equation}

\section{Results}\label{sec:results}

In the following, we summarize key TRPMD results for silicon, focusing on the mean squared displacement, the ring-polymer radius of gyration, the velocity autocorrelation function, and the resulting temperature-dependent EELS spectra. Details of the TPRMD implementation and EELS spectra calculations can be found in the Appendix~\ref{sec:TRPMDTACAWmethod}.

\begin{figure*}[t!]
  \centering
  \includegraphics[width=1.\textwidth]{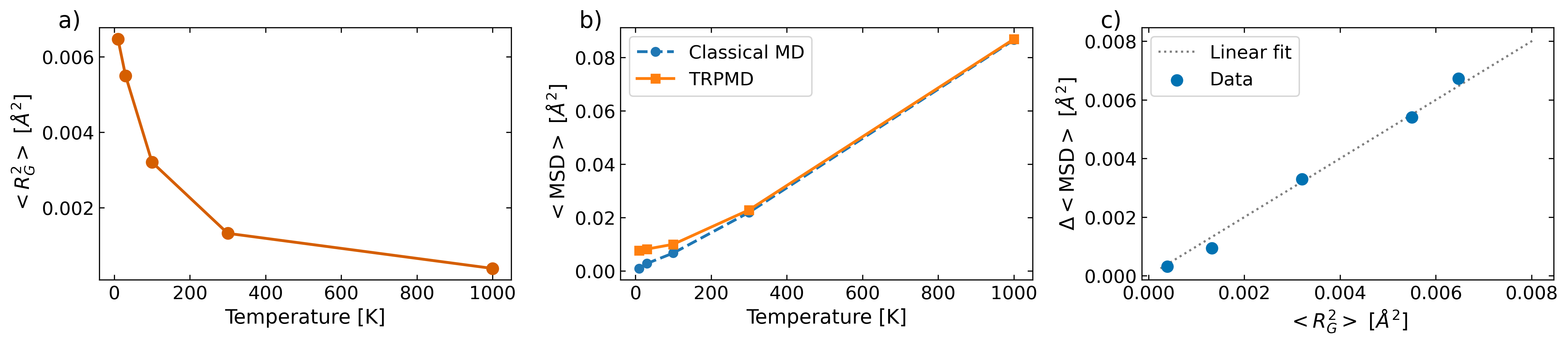}
  \caption{Numerical simulation of TRPMD for silicon were performed at several temperatures. After the thermalization, the emsamble average square of radius of gyration of ring polymer obtained from TPRMD is shown in panel (a). The ensemble averaged MSD computed evaluated after thermalization from classical MD and TRPMD simulations, are shown in the panel (b). In the panel (c), the ensemble average MSD against squared of radius of gyration is plotted, together with a linear fit. }
  \label{fig:r2_MSD_temps}
\end{figure*}

\subsection{Quantum delocalization and equilibrium dynamical observables}

TRPMD simulations were performed for silicon over a temperature range from 10 K to 1000 K. As discussed above in Section~\ref{subsec:ObsTRPMD}, one of the key dynamical observables accessible within the TRPMD framework is the radius of gyration of the ring polymer, which quantifies the spatial delocalization of the nuclear paths and provides a direct measure of nuclear quantum effects, see Fig \ref{fig:r2_MSD_temps}a. From the figure, we observe that at $T=1000 \text{ K}$ the radius of gyration is significantly reduced compared to lower temperatures, indicating that the ring polymer collapses toward its centroid and the nuclear dynamics approaches the classical limit. In contrast, at low temperatures, the larger radius of gyration reflects enhanced quantum delocalization arising from zero-point motion.

To systematically compare the differences between TRPMD and classical MD, the MSD is averaged over the equilibrated portion of each trajectory at each temperature and is used to construct the temperature-dependent comparison; see Fig.~\ref{fig:r2_MSD_temps}b. At $T=1000 \text{ K}$, the MSD values from TRPMD and MD are nearly identical, indicating that nuclear quantum effects are negligible and that the system behaves classically. As the temperature decreases, the deviation between TRPMD and MD increases, reflecting the growing importance of nuclear quantum effects below $\sim 200 \text{ K}$.

Finally, a linear relation is observed between the difference of classical vs TRPMD values of MSDs and the squared radius of gyration, Fig.~\ref{fig:r2_MSD_temps}c. This behavior indicates that the squared radius of gyration serves as a quantitative measure of the delocalization arising from nuclear quantum fluctuations and characterizes the deviation between TRPMD and classical molecular dynamics.

\begin{figure}[!htp]
  \centering
  \includegraphics[width=0.48\textwidth]{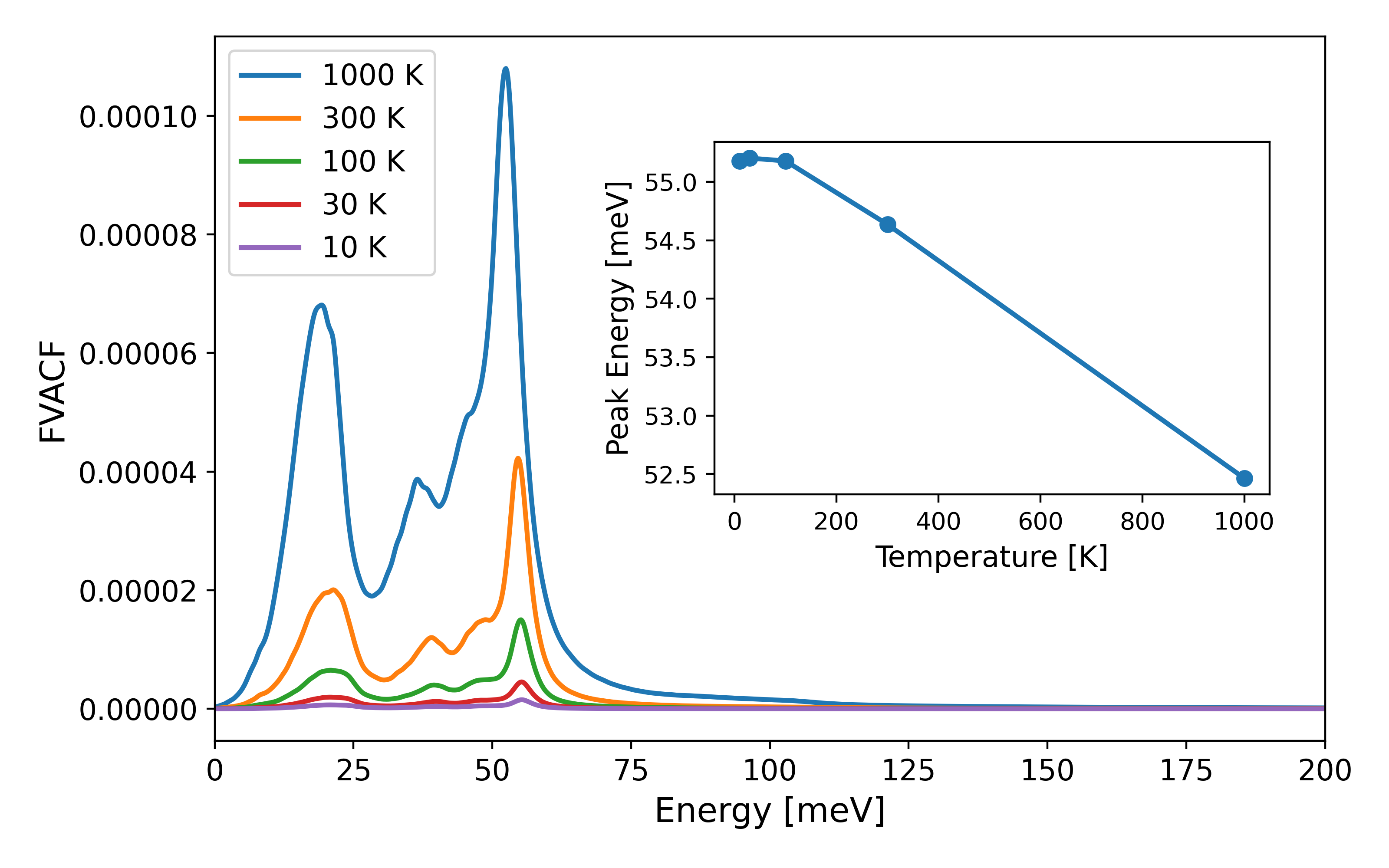}
  \caption{The FVACF for silicon at various temperatures is computed from centroid velocity of ring-polymer. The inset shows the peak energy of the FVACF spectrum as a function of temperature.}
  \label{fig:rpmd_facf}
\end{figure}

The vibrational density of states can be characterized by the Fourier transform of the VACF. Here we compute this quantity from TRPMD trajectories using its centroid velocity, and the resulting spectra for silicon at different temperatures are shown in Fig.~\ref{fig:rpmd_facf}.  With increasing temperature, the spectral features broaden, and the overall intensity increases, consistent with enhanced thermal motion. By integrating the FVACF, we observe that the integral increases approximately linearly with temperature, with a slope of about $2.5 \times 10^{-6} \mathrm{ meV} \mathrm{~K}^{-1}$ and a small intercept on the order of $10^{-8} \mathrm{ meV}$. Notably, the energy of the dominant spectral peak exhibits an apparent blueshift upon cooling, indicating an effective stiffening of the corresponding vibrational mode due to reduced anharmonic softening at low temperatures, see inset in Fig.~\ref{fig:rpmd_facf}.

\subsection{Temperature dependent angle-resolved phonon EELS}

\begin{figure*}[!htp]
  \centering
  \includegraphics[width=1.\linewidth]{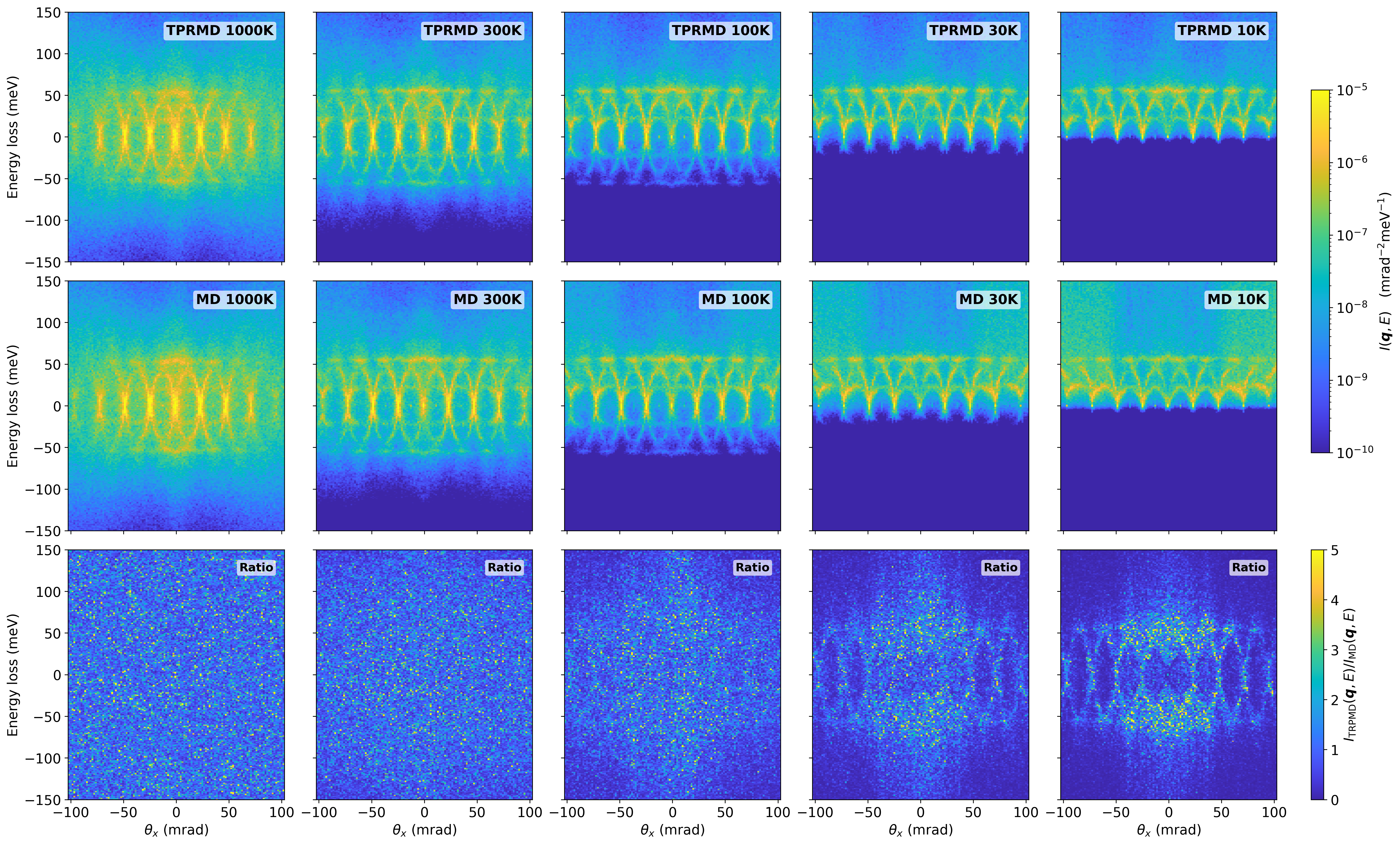}
  \caption{The temperature dependent angle resolved phonon EELS of silicon were obtained from the scattering intensity $I(q_x,q_y=0,E)$ is computed using TRPMD and classical molecular dynamics. The calculations span the temperature range from 1000 K down to 10 K. The TRPMD and classical results are shown in the first and the second row, respectively. The bottom row displays the ratio between the two angle-resolved spectra, highlighting the temperature-dependent deviations arising from nuclear quantum effects in low temperature limit.}
  \label{fig:Phonon_Dispersion}
\end{figure*}

The temperature-dependent angle-resolved phonon EELS are summarized in Fig.~\ref{fig:Phonon_Dispersion}. At $1000$ and $300~\mathrm{K}$, both TRPMD and classical MD show pronounced thermal broadening, as expected from stronger anharmonic scattering and shorter phonon lifetimes. Particularly, at 1000 K, the spectra are nearly symmetric about the elastic line because thermally populated vibrational states contribute on both the energy-loss and energy-gain sides. The ratio maps are correspondingly noise-like, indicating that classical dynamics provide an adequate description in this regime.

Below $300~\mathrm{K}$, the spectra evolve in accordance with detailed balance. The energy-gain intensity is progressively suppressed as the vibrational population decreases, and the branches sharpen as thermal disorder is reduced. In the cryogenic regime, however, the TRPMD and classical spectra no longer coincide. The nonzero residual intensity in the ratio maps at $30$ and $10~\mathrm{K}$ shows that the quantum and classical descriptions redistribute spectral weight differently, even when both capture the overall approach to the spectral structure. These deviations are attributed to nuclear quantum effects, which are absent in classical MD.

\subsection{Temperature-dependent EELS spectra}

\begin{figure*}[!htp]
  \centering
  \includegraphics[width=1.\textwidth]{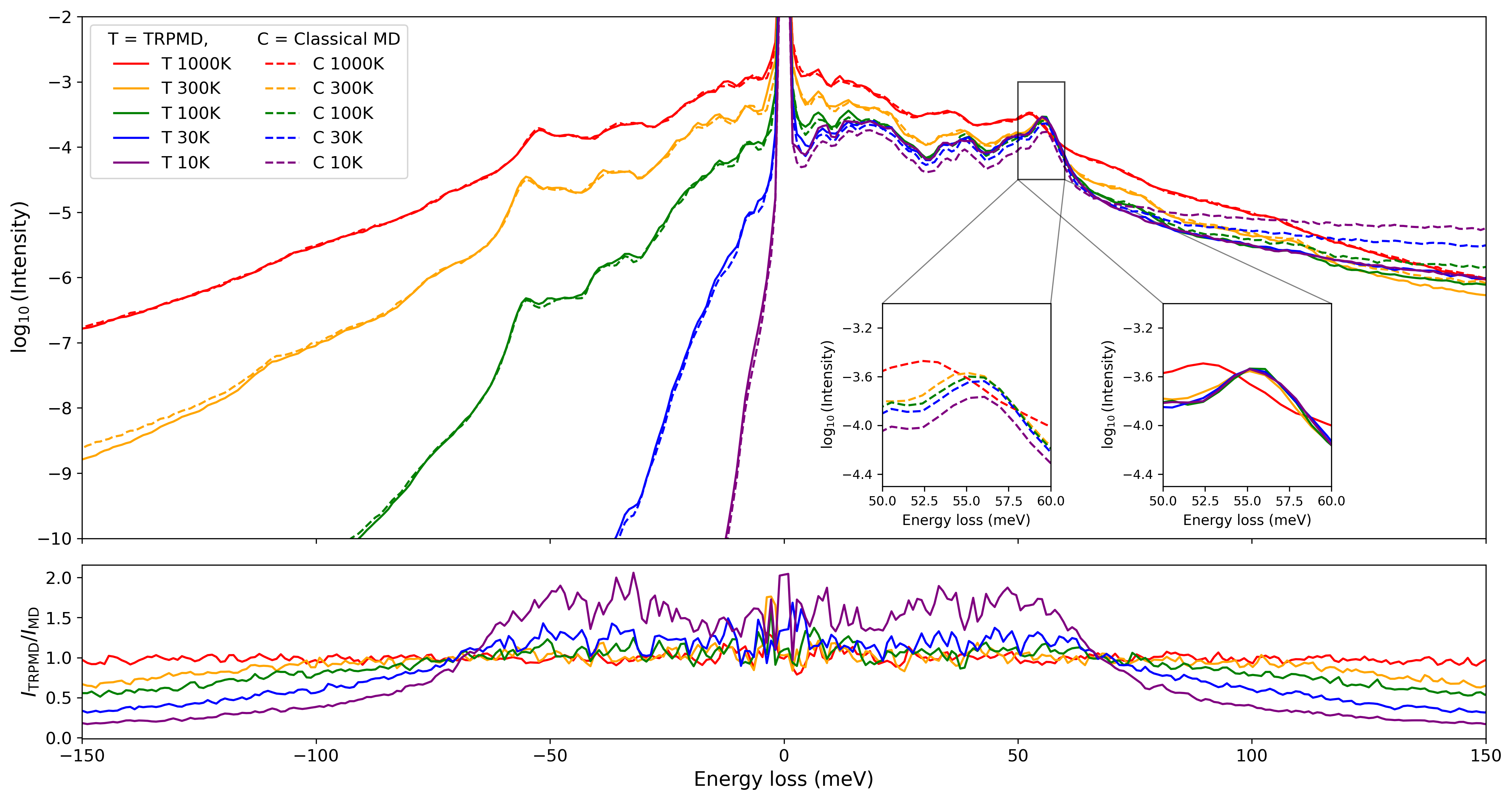}
  \caption{The EELS spectra at different temperatures were obtained from the scattering intensity $I(\boldsymbol{q}, E)$ on a logarithmic scale by placing the detector at the center with a collection semi-angle of 35 mrad. The first panel shows the EELS spectra for different temperatures computed using TRPMD and classical MD, represented by solid and dashed lines, respectively. The second panel displays the difference ratio the two spectra.}
  \label{fig:EELS 35mrad}

\end{figure*}

Electron energy-loss spectroscopy (EELS) in the scanning transmission electron microscope provides access to vibrational excitations through the energy distribution of inelastically scattered electrons. In the present context, the central quantity of interest is the temperature-dependent vibrational EELS signal predicted by the TRPMD-TACAW framework, see Fig.~\ref{fig:EELS 35mrad}. The simulations compute the EELS signal from $I(\boldsymbol{q}, E)$ by placing the detector at the center of the diffraction pattern with a collection semi-angle of 35 mrad. The spectra are obtained using both TRPMD and classical molecular dynamics over a temperature range from 1000 K down to 10 K, as shown in the top panel. The bottom panel displays the relative difference in intensity between the two approaches.

In the difference panel, one observes that at high temperatures, the two methods produce nearly identical results. As the temperature decreases, however, a clear difference in intensity emerges between TRPMD and classical molecular dynamics. More specifically, in the energy window from approximately -70 meV to 70 meV, the TRPMD intensity becomes systematically larger than that obtained from classical MD at low temperatures. This enhancement reflects the contribution of nuclear quantum effects, in particular, the presence of zero-point motion.

Additional differences between the two methods appear in the optical phonon region. In classical MD, the optical peak decreases in intensity as the temperature is reduced and simultaneously exhibits a blue shift upon cooling. By contrast, the TRPMD spectra show a weaker temperature dependence around the phonon peak. The position of the optical phonon peak remains essentially unchanged from 300 K down to 10 K, with no evident blue shift. A noticeable shift is only present when comparing the high-temperature spectrum at 1000 K with that at 300 K.

The change in the phonon peak energy as the temperature varies originates from a combination of lattice thermal expansion and anharmonic phonon scattering. At high temperatures, these effects are significant, leading to noticeable temperature-dependent shifts in the phonon frequencies. In the cryogenic regime, the lattice dynamics approaches the harmonic limit, where anharmonic scattering is strongly suppressed and phonon energies become nearly temperature independent. Within the first Born approximation, for single-phonon excitation, the energy-loss intensity is proportional to $n(E, T)+1$, where $n(E, T)$ is the Bose-Einstein distribution \cite{nicholls_theory_2019}. In the regime where $E \gg k_B T, n(E, T)$ becomes small, the phonon peak intensity is therefore effectively temperature independent. At room temperature $T =300$ K, we have $k_B T \approx 26 \mathrm{meV}$. Consequently, for optical phonon energies well above this scale, the phonon peak intensity remains essentially temperature independent over the range from 10 K to 300 K. At temperatures of 1000 K, this approximation is no longer valid, as $k_B T$ exceeds the optical phonon energy. This behavior is captured by TRPMD, indicating that this method correctly reproduces near-harmonic lattice dynamics while retaining the quantum nuclear fluctuations absent in classical molecular dynamics. Beyond the optical phonon peak, multi-phonon excitations and anharmonicity become relevant, leading to additional contributions to the energy-loss spectrum.

\section{Conclusion}\label{sec:conclusion}

We have formulated a TRPMD-TACAW framework for temperature-dependent vibrational EELS that combines the TACAW method with a path-integral description of quantum nuclei. The method preserves the key advantages of TACAW, including dynamical diffraction and multiple scattering, while extending the nuclear dynamics beyond the classical approximation.

For crystalline silicon, the calculations show that classical molecular dynamics and TRPMD descriptions converge at high temperatures but diverge systematically in the cryogenic regime. The deviations appear consistently in ring-polymer observables, angle resolved phonon EELS, and the EELS spectra, indicating that nuclear quantum effects quantitatively influence the vibrational spectra. The present framework therefore provides a practical starting point for the quantitative interpretation of low-temperature vibrational EELS in modern cryogenic transmission electron microscopy.

\appendix

\section{Setting up the TRPMD-TACAW}\label{sec:TRPMDTACAWmethod}

The unit cell of silicon is constructed with the $z$-axis aligned along the $[110]$ direction. First, we begin by determining the lattice constant for silicon using a 5x7x7 supercell of cubic unit cells. The lattice constant is obtained as an average lattice parameter in an NPT simulation at a temperature $T=300K$ and zero pressure.

The TRPMD simulations were performed within the \texttt{i-PI} framework \cite{litman2024pi}. In this workflow, \texttt{i-PI} acts as the path-integral dynamics driver and communicates with \texttt{LAMMPS} \cite{LAMMPS_paper_2022}, which serves as the force engine for the underlying interatomic potential through the client–server interface. To set up the TRPMD simulations, we consider silicon with a 5x7x52 supercell of cubic unit cells. The machine-learning SNAP interatomic potential for silicon \cite{thompson_interatomic_2015,zuo_performance_2020} was employed. The simulations were performed using an integration time step of 0.5 fs. Observables were sampled every 5 steps, corresponding to a sampling interval of $\Delta t=2.5 \mathrm{fs}$, over a total of 20000 integration steps. After an initial thermalization of 5000 integration steps, only the equilibrated portion of the trajectories was used for subsequent calculations.

To assess the ergodicity of the trajectories, different thermostat schemes were tested, including the Langevin thermostat and the PILE-L and PILE-G thermostats, with their respective damping parameters, see Appendix. \ref{sec:diffThermostat}. For standard molecular dynamics $(n=1)$, a Langevin thermostat was employed. For TRPMD simulations $(n>1)$, the PILE-L thermostat was used. For both thermostats, we used a thermostat relaxation time of 200 fs. Simulations were performed at temperatures of 10~K, 30~K, 100~K, 300~K, and 1000~K for silicon. At each temperature, the number of beads was chosen as $n=165$, 60, 20, 9 and 4, respectively, in TRPMD in accordance with Eq.~\eqref{eq:boundbeads}.

The exit wavefunctions were computed using \texttt{pyms}, a Python package implementing the multislice method \cite{brown2020python}. To compute the exit wavefunctions, the incident electron beam energy was set to 60 keV. Wave propagation was performed on a numerical real-space grid of $400 \times 399$ points, while the full thickness of the simulation supercell was discretized into 208 slices along the beam-propagation direction.

After discarding the initial thermalization segment, the remaining trajectory is analyzed using a sliding window (chunked) time-series approach to evaluate the scattering intensity $I(\boldsymbol{q}, E)$ \cite{welch1967use}. In order to evaluate $I(\boldsymbol{q}, E)$, the following parameters are introduced:
\begin{itemize}
  \item $N$: the total number of stored trajectory frames per bead;
  \item $C$: the chunk length, corresponding to the number of time samples per chunk;
  \item $s$: the stride between sampled frames within a chunk with $s>1$; When $s=1$, we use every stored frame, which maximizes time resolution, thereby incurring the highest computational cost. When choosing $s>1$, it downsamples in time. It reduces cost and high-frequency noise but limits maximum resolvable energy.
  \item $n_{\text{therm}}$: the number of initial frames discarded as thermalization;
  \item $o$: the offset between successive chunk starting points.
\end{itemize}

Each chunk starts at the frame index
\begin{equation}
n^{(m)}=n_{\text {therm }}+m o, \quad m=0,1,2, \ldots
\end{equation}
subject to the constraint
\begin{equation}
n^{(m)}+(C-1) s<N \text {. }
\end{equation}
The total number of chunks is therefore
\begin{equation}
N_{\text {chunk }}=\left\lfloor\frac{N-C s-n_{\text {therm }}}{o}\right\rfloor+1 .
\end{equation}
Within a given chunk $m$, the sampled frame indices are
\begin{equation}
n_i^{(m)}=n^{(m)}+i s, \quad i=0, \ldots, C-1 .
\end{equation}

Let $\mathbf{R}^{(k)}\left(n_i^{(m)}\right)=\left\{\mathbf{R}_a^{(k)}\left(n_i^{(m)}\right)\right\}_{a=1}^{N_{\text {atom }}}$ denote the atomic configuration of bead $k=1, \ldots, n$ at frame $n_i$. For each bead and sampled frame, this configuration is used as input to a multislice calculation, yielding a momentum-space exit wave
\begin{equation}
\psi_{m, i}^{(k)}(\boldsymbol{q})=\mathcal{M}\left[\mathbf{R}^{(k)}\left(n_i^{(m)}\right)\right] \psi_0
\end{equation}
where $\psi_0$ is the incident plane-wave probe, and $\mathcal{M}$ denotes the multislice propagation operator.

Within each chunk, the exit-wave time series is multiplied by a Hann window $w_i$, given by
\begin{equation}
    w_i=\frac{1}{2}\left(1-\cos \frac{2 \pi i}{C}\right), \quad i=0,1, \ldots, C-1
\end{equation}

The windowed exit waves are then averaged over the beads according to
\begin{equation}
\bar{\psi}_{m, i}(\boldsymbol{q})=\frac{1}{n} \sum_{k=1}^n w_i \psi_{m, i}^{(k)}(\boldsymbol{q}), \quad i=0, \ldots, C-1
\end{equation}
A discrete Fourier transform is applied along the chunk-time index,
\begin{equation}
\Psi_{m, \ell}(\boldsymbol{q})=\sum_{j=0}^{C-1} \bar{\psi}_{m, i}(\boldsymbol{q}) e^{-2 \pi i \ell j / C}, \quad \ell=0, \ldots, C-1
\end{equation}
The corresponding intensity is defined as $I_{m, \ell}(\boldsymbol{q})=\left|\Psi_{m, \ell}(\boldsymbol{q})\right|^2$. Finally, the scattering intensity is obtained by averaging over all chunks,
\begin{equation}
I_{\ell}(\boldsymbol{q})=\frac{1}{N_{\text {chunk }} C} \sum_{m=0}^{N_{\text {chunk }}-1} I_{m, \ell}(\boldsymbol{q}) .
\end{equation}

The choice of the chunk length and offset determines the balance between spectral resolution and statistical convergence of the computed intensity $I_{\ell}(\boldsymbol{q})$. In our simulations, we tested different values of the chunk length, stride, and offset, and observed no significant change in the resulting spectra for silicon. We therefore set the chunk length to 600 time steps, the stride to 3 time steps, and the offset equal to the chunk length.

In general, the chunk length sets the effective time window over which temporal Fourier transforms are performed and therefore controls the energy resolution, $\Delta E \sim 2 \pi \hbar /(C s)$. Increasing chunk length improves frequency resolution but simultaneously reduces the number of available chunks, leading to increased statistical noise. Moreover, when $C s$ significantly exceeds the intrinsic correlation time of the nuclear dynamics, the signal within a chunk becomes weakly correlated.

The offset determines the degree of overlap between successive chunks and hence the number of statistically independent samples contributing to the final average. Within the framework of Welch’s method \cite{welch1967use}, small offsets produce strongly overlapping chunks that improve variance reduction but introduce correlations between samples, whereas large offsets yield nearly independent chunks at the cost of reduced averaging. In practice, choosing the offset to be a fraction of the chunk duration provides a favorable compromise between statistical efficiency and sample independence.

\section{Spurious resonance in EELS}\label{sec:diffThermostat}

\begin{figure}
    \centering
    \includegraphics[width=1.\linewidth]{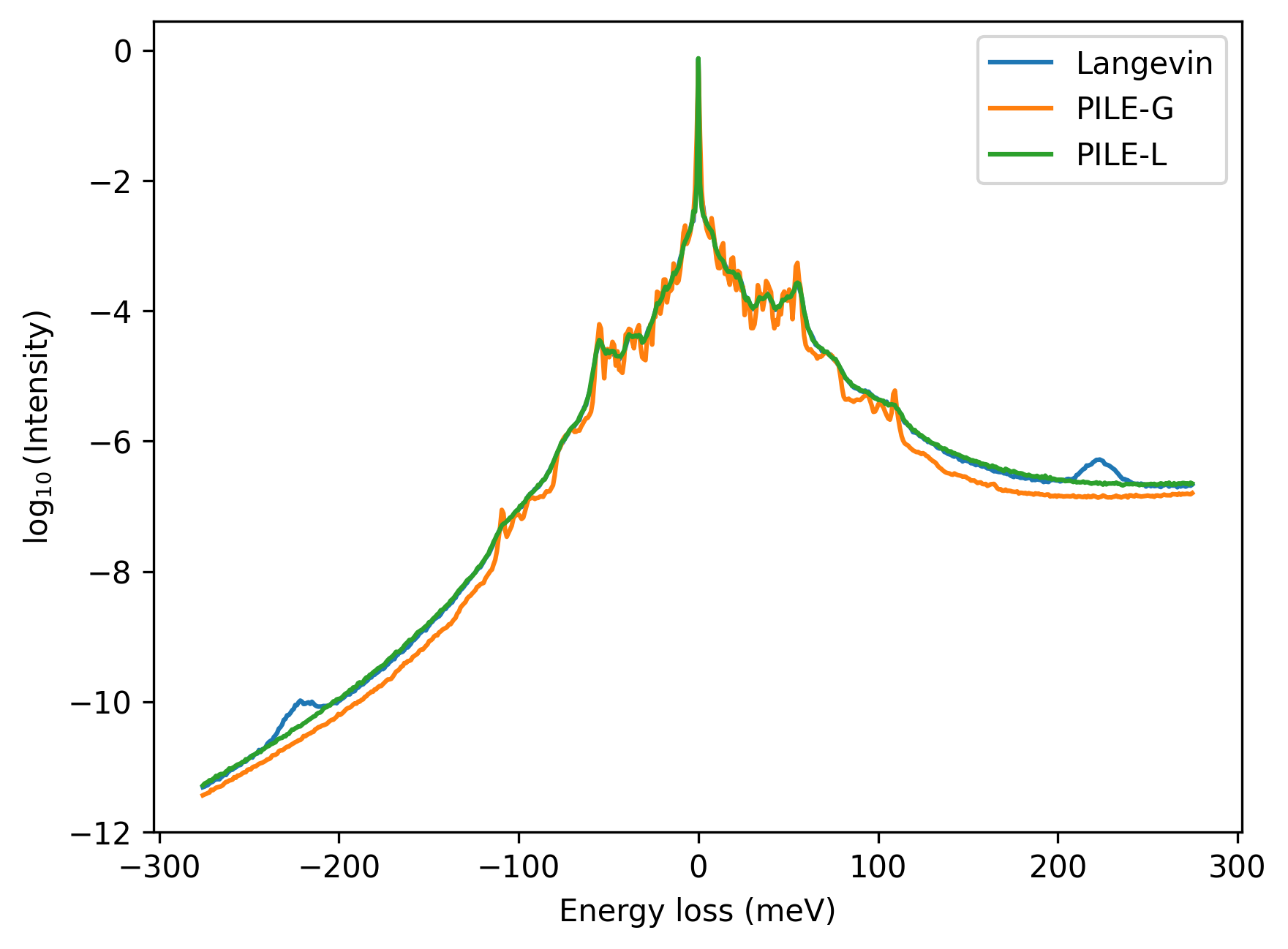}
    \caption{The EELS spectra at temperature 300~K obtained from TRPMD using three different thermostats: Langevin, PILE-L and PILE-G, with thermostat relaxation time set to 200~fs. The scattering intensity $I(\boldsymbol{q}, E)$ is plotted on a logarithmic scale. The same detector as in Fig.~\ref{fig:EELS 35mrad} was assumed: collection semi-angle of 35~mrad around the center of diffraction pattern.}
    \label{fig:resonance_PILE_Langevin}
\end{figure}

The EELS from $I(\boldsymbol{q}, E)$, is computed at $T=300 \mathrm{~K}$ for silicon using TRPMD by placing the detector at the center with a collection semi-angle of 35 mrad; see Fig. \ref{fig:resonance_PILE_Langevin}. Different types of thermostats are employed in TRPMD, namely standard Langevin, PILE-L, and PILE-G, while maintaining the same thermostat relaxation time of 200 fs, which results in different EELS. We observe that the use of the standard Langevin thermostat introduces a spurious resonance around $\pm 220 \mathrm{meV}$. In contrast, both PILE-L and PILE-G suppress this unphysical feature. Among these, PILE-L yields a smoother spectrum with reduced noise. Consequently, for subsequent TRPMD simulations at other temperatures, we adopt the PILE-L thermostat with a relaxation time of 200 fs .


%
\bibliographystyle{apsrev4-2}

\bibliography{references}


\end{document}